\documentclass[12pt,a4paper]{article}
\usepackage[utf8]{inputenc}
\usepackage[T1]{fontenc}
\usepackage{amsmath,amssymb,bbm}
\usepackage{graphicx}

\begin{document}
\textwidth=135mm
 \textheight=200mm
\begin{center}
{\bfseries Composite Fermions in Medium: Extending the Lipkin Model
%\footnote{{\small Poster at the 32th Max-Born-Symposium and HECOLS workshop on 
%"Three Days of Phase Transitions in Compact Stars, Heavy-Ion Collisions and Supernovae", 
%Institute for Theoretical Physics, University of Wroc\l{}aw, Wroc\l{}aw, Poland, 
%February 17--19, 2014.}}
}
\vskip 5mm
S. Liebing$^*$\footnote{Email: liebings@physik.tu-freiberg.de} and 
D. Blaschke$^{\dag,\ddag}$
\vskip 5mm
{\small {\it $^*$ Institute of Theoretical Physics, TU Bergakademie Freiberg, 
%09596 
Freiberg, Germany}}\\
{\small {\it $\dag$ Uniwersytet Wroc\l{}awski, Instytut Fizyki Teoretycznej,
%pl. M. Borna 9, 50-204 
Wroc\l{}aw, Poland}}\\
{\small {\it $^\ddag$ Joint Institute for
Nuclear Research, 141980 Dubna, Russia}} \\
\end{center}
\vskip 5mm
\centerline{\bf Abstract}
The role of phase space occupation effects for the formation of two- and 
three-particle bound states in a dense medium is investigated within an 
algebraic approach suitable for systems with short-range interactions.
It is shown that for two-fermion bound states due to the account of the 
exchange symmetry (phase space occupation) effect (Pauli blocking) in a dense 
medium the binding energy is reduced and vanishes at a critical density 
(Mott effect).
For three-fermion bound states, within a Faddeev equation approach, the 
intermediate formation of pair correlations leads to the representation as 
a suitably symmetrized fermion-boson bound state. 
It is shown that the Bose enhancement of fermion pairs can partially 
compensate the Pauli blocking between the fermions.
This leads to the general result obtained by algebraic methods:
three-fermion bound states in a medium with high phase space occupation
appear necessarily as Borromean states beyond the Mott density of the 
two-fermion bound state.
\vskip 10mm

\section{\label{sec:intro}Introduction}

From the book of Lipkin \cite{Lipkin} we know that in contrast to elementary 
bosons the deuterons (and other two-fermion bound states) being composite 
bosons localized in space require a Fourier spectrum of plane waves 
describing their nucleonic constituents which blocks a part of the fermionic
phase space according to the Pauli principle. 
Contrary to naive expectation this leads to a statistical repulsion between 
bosons at high densities.
This contribution discusses the question: what changes when a system of 
composite fermions is considered which consists of a boson-fermion pair where
the boson is composed of two fermions?
We provide an argument based on the algebraic treatment of these bound states 
that composite fermions are likely to form Borromean states \cite{Zhukov:1993aw}
in a dense medium,
since Pauli blocking shall make the fermion pair an unbound correlation 
with a Bose enhancement factor which partly compensates for the Pauli blocking
of the valence fermion so that the composite fermion remains a bound state. 

We investigate bound states in many-particle systems and their stability under
conditions of high phase space occupation. 
This situation is particularly prominent at low temperatures where we face a
competition between Bose-Einstein condensation of bosonic 
two-particle bound states and BCS condensation of Cooper pairs. 
The point there is that the Mott effect due to Pauli blocking drives 
the binding energy to zero. 
Nevertheless, there is a strong resonant two-particle correlation in the scattering 
state spectrum above the continuum edge.
This is general for two-particle correlations in Fermi systems, e.g., excitons in an 
electron-hole plasma \cite{Stolz}, hydrogen in a non-ideal electron-proton plasma 
\cite{Kremp}, deuterons in nuclear matter \cite{Schmidt}, mesons and diquarks in 
quark matter \cite{David1,Redmer}. 
%(See also Book "Metal-to-Nonmetal Transitions'' Springer (2010)\cite{Redmer}).

This work considers also the question of three-particle bound states in a medium,
examples being superfluid $^3$He,  Tritons/$^3$He in nuclear matter (sun, stars) 
and nucleons in quark matter.
One important point there is also the role of Borromean states were the 
two-particle system is already unbound while the three-particle state remains
bound at the same time.
Here we will study the specific case that the two-particle state is bound when
isolated but can undergo density ionization/dissociation (Mott effect)
in a medium with high phase space occupation due to Pauli blocking. 
The question arises what this entails for the three-particle bound state.
We demonstrate under very general conditions (basically only that the system 
is bound by short range interactions such as nuclear force) by using algebraic 
methods that the three-particle bound state has necessarily to persist as a 
bound state at densities exceeding the Mott density of the two-particle state.
This effect of an ``in-medium Borromean state'' will be the stronger the 
stronger the pairing correlation in the three-particle state is.

\section{\label{sec:comBo}Composite Bosons}
\begin{figure}[hbt]
   \centering
   \includegraphics[width=0.7\textwidth,keepaspectratio=true]{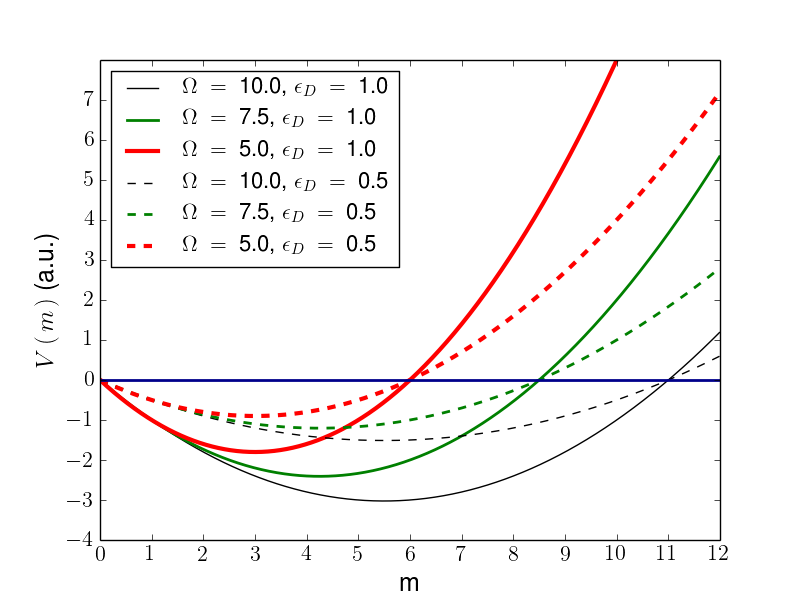}
     \caption{The figure shows how Bose enhancement evolves with growing 
particle density and gets overcompensated by the Pauli blocking. 
Here $\Omega$ is the spatial volume and  $\epsilon_D$ the eigenvalue of the 
interaction operator $V$.}
   \label{fig:spacefill}
 \end{figure}

In the beginning we consider elementary fermions coupled together to form 
composite bosons. 
The creation operator for such a boson resembles the subsequent application of two 
fermion creation operators. 
This can be represented as a linear combination of two plane wave functions 
(\ref{eqn:pwa}). 
Thereby $r_i$ are the coordinates of the nucleons, $R$ is the center of mass 
and $r$ the relative coordinate of the composite particle (deuteron). 
In the same way we define the boson creation operator $D^\dagger_{2K}$ and 
annihilation operator $D_{2K'}$ in momentum space.
\small
 \begin{align}
      <R,r|D,2K> = e^{2 i K R}~\phi(r) = 
\sum\limits_q g_q~ e^{i\left(K+q\right)r_1}~ e^{i\left(K-q\right)r_2} ~,
\label{eqn:pwa}\\
	 D^\dagger_{2K} = 
\sum\limits_q g_q ~a^\dagger_{\left(K+q\right)\uparrow}~ 
a^\dagger_{\left(K-q\right)\downarrow} ~,~
	 D_{2K'} = 
\sum\limits_q g_q ~a_{\left(K'-q\right)\downarrow}~a_{\left(K'+q\right)\uparrow}~. 
\nonumber
\end{align}
\normalsize
The next step in evaluating the properties of the created boson is to 
calculate the commutator between creation and annihilation operator. 
In contrast to elementary bosons the commutator does not vanish
\small 
 \begin{align}
      \left[D_{2K'},D^\dagger_{2K}\right] =& \,\delta_{KK'}-\Delta_{KK'}~,\\
       \Delta_{KK'} =& 
\sum\limits_q g_q~ \{ g_{\left(K'-K+q\right)}~
a^\dagger_{\left(2K-K'-q\right)\downarrow}~a_{\left(K'-q\right)\downarrow} 
       \notag\\&+ g_{\left(K'-K-q\right)}~ 
a^\dagger_{\left(2K-K'+q\right)\uparrow}~ a_{\left(K'+q\right) \uparrow} \}\\
	\Delta_{KK} =& \sum\limits_q~ g^2_{K-q} ~n_{q\downarrow} + g^2_{q-K}~ n_{q\uparrow}~.
  \end{align}
\normalsize
The violation of the elementary Bose commutator is proportional to the 
occupation numbers of the fermionic constituents.
For investigating phase space occupation effects in a system of $m$ composite
bosons we apply a quasispin formalism. 
After doing so we apply a simple interaction operator $V = - \epsilon_D N_D$ 
to the multiparticle state just created. 
Thereby $N_D$ is the boson number operator, and the strength of the 
two-particle interaction is $G=\frac{-\epsilon_D}{\Omega}$, with $\Omega$ 
being the volume of the phase space and $-\epsilon_D$ being an eigenvalue of 
the interaction operator $V$ 
\small
\begin{align}
    V&\left(D^\dagger_{2K}\right)^m\left\lvert0\right\rangle \nonumber\\
    &= -\left(\frac{\epsilon_D}{\Omega}\right) \left(\frac{\Omega}{2}\left(
    \frac{\Omega}{2}+1\right)-\left(\frac{2m-\Omega}{2}\right)^2 +\frac{2m-\Omega}{2}\right)
    \left(D^\dagger_{2K}\right)^m\left\lvert0\right\rangle
    \\ &= - m~\epsilon_D\left(1-\frac{m-1}{\Omega}\right)\left(D^\dagger_{2K}\right)^m
    \left\lvert0\right\rangle~. 
\label{eqn:mulbos}
\end{align}
\normalsize
Equation (\ref{eqn:mulbos}) can be evaluated for different values of $\Omega$ 
and $\epsilon_D$. The result is shown in figure \ref{fig:spacefill}.
For low particle densities the binding energy increases due to Bose 
enhancement. By adding more particles the phase space gets filled up and it 
starts to cost energy to add more particles. 
This is due to the Pauli blocking of the fermionic substructure.
  
\section{\label{sec:CoBa}Composite Fermions (Baryons) }
Similar to the algebraic treatment of composite bosons we can also consider 
baryons as composite fermions represented by creation and annihilation 
operators with one fermion with one boson operator which stand for the 
operators of quarks and diquarks, respectively
\small 
 \begin{align}
%       <R,r|N,K> = 
      N_K^\dagger = \sum_q g_q ~ a^\dagger_{\frac{K}{2}+q} b^\dagger_{\frac{K}{2}-q} ~,~
% 	N_K^\dagger = \sum_q g_q a^\dagger_{\frac{K}{2}+q} b^\dagger_{\frac{K}{2}-q} \\
 	N_{K'} = \sum_q g_q ~ b_{\frac{K'}{2}-q} a_{\frac{K'}{2}+q }~.
  \nonumber
 \end{align}
\normalsize
For these composite operators we calculate the anticommutator between 
composite fermions which, similar to the case of the composite bosons receives
a correction term beyond the Kronecker symbol
\small
\begin{align}
        [N_{K'},N_K^\dagger]
=& \sum\limits_q g_q \left( b_{\frac{K'}{2}-q} a_{\frac{K'}{2}+q} N^\dagger_K 
+ N^\dagger_K b_{\frac{K'}{2}-q} a_{\frac{K'}{2}+q} \right)  \\
=&~\delta_{K,K'} + \Delta_{K,K'}~,\\
\Delta_{K,K'} =& \sum\limits_q g_q  
\left( g_{\frac{K-K'}{2}-q} b^\dagger_{K-\frac{K'}{2}-q} b_{\frac{K'}{2}-q} 
- g_{\frac{K-K'}{2}+q} a^\dagger_{K-\frac{K'}{2}+q} a_{\frac{K'}{2}+q}  \right)~.
\end{align}
\normalsize
For equal moments $K' \rightarrow K$ this correction term takes the form
\small
\begin{align}
  \Delta_{K,K} 	&= \sum\limits_q g^2_q  \left(   n^{(D)}_{\frac{K}{2}-q} 
	- n^{(Q)}_{\frac{K}{2}+q}  \right)~,
\end{align}
\normalsize
which makes the different phase space occupation effects for bosonic (D) and
fermionic (Q) constituents apparent.
In contrast to the composite boson case we observe competing phase space 
occupation effects: the Pauli blocking of quarks can be partially compensated
by Bose enhancement of diquarks.
This agrees with recent calculations of baryon properties in the PNJL model 
\cite{Blanqier, David2}.
%, were the phase space occupation factors show partial 
%compensation of Pauli blocking and Bose enhancement. 
In particular, baryons appear to be bound under conditions when diquarks are 
already unbound due to Pauli blocking (in-medium borromean state). 

Two fermions are not allowed to be in the same quantum state therefore it is 
not correct to go to the limit $K' \rightarrow K$\,.
%As we are talking about the formation of fermions it is not correct to go to 
%the limit $K' \rightarrow K$, because two fermions are not allowed to be in 
%the same quantum state. 
This was only done in order to investigate the effects qualitatively.
For a systematic study of the properties of composite fermions or multi-fermion
states it is required to deal with the case $K\neq K'$. 
Then it is not immediately possible to express $\Delta_{K'K}$ in 
terms of densities. 

To quantify the effect it is required to develop a method to investigate the 
interactions between composite particles. 
%This can be realized in the way it was done for the multiple bosons in a kind 
%of quasi spin formalism, see \cite{blaizot}. 
%It is also possible to simulate the effect by Hartree-Fock or more 
%sophisticated density functional treatments. 
To this end cluster virial expansion techniques will be developed for the case 
of nonideal plasmas with multiparticle correlations \cite{Ropke:2012qv}.
This goes beyond the scope of the present work and will be addressed in a forthcoming 
publication.
 
\section{\label{sec:Con}Conclusions}
The algebraic treatment of a system of many composite bosons by Lipkin was 
reproduced and the results were visualized. 
Additionally, the approach was applied to describe composite fermions. 
We introduced creation and annihilation operators for a boson-fermion pair. 
The calculation of their anticommutator shows two competing density effects 
identified as Bose enhancement and Pauli blocking for the composite fermion. 
Consequently, there is a partial compensation which stabilizes the composite fermion
in the medium. 
This can lead to the existence of Borromean states where the two-particle 
state is unbound while the three-particle state stays bound. 
Such kind of behavior was already discussed within the PNJL model \cite{David2}. 
Quantifying this effect in the presented approach requires the calculation of 
multi-fermion states, which goes beyond the focus of the current work. 

%\newpage
\vskip 10mm
\centerline{\bf Acknowledgment}
This work was supported in part by the Heisenberg-Landau Programme,
% of the BMBF,
%Helmholtz Association within the "HISS Dubna" programme, 
by the Polish NCN under grant no. UMO-2011/02/A/ST2/00306
and by the COST Action MP1304 
%`` Exploring fundamental physics with compact stars (NewCompStar)''
"NewCompStar". 
%S.L. thanks the Bogoliubov Laboratory for Theoretical Physics at JINR Dubna 
%and University of Wroclaw for hospitality during several
%collaboration visits.
%D.B. received partial support from the Polish MNiSW under grant no. 1009/S/IFT/14.

\end{document}